\documentclass[reprint,aps,prl,twocolumn,superscriptaddress,showpacs,amsmath,amssymb]{revtex4-1}
\usepackage{amsmath}
\usepackage{amssymb}
\usepackage[dvipdfmx]{graphicx}
\usepackage{here}
\usepackage{bm}
\usepackage{color}

\begin{document}

\title{
First-Principles Design of a Half-Filled Flat Band of the Kagome Lattice \\ in Two-Dimensional Metal-Organic Frameworks
}
\author{Masahiko G. Yamada}
%\email[E-mail me at: ]{m.yamada@issp.u-tokyo.ac.jp}
\affiliation{Institute for Solid State Physics, University of Tokyo, Kashiwa 277-8581, Japan}
\affiliation{Department of Physics, University of Tokyo, Hongo, Tokyo 113-0033, Japan}
\author{Tomohiro Soejima}
\affiliation{Department of Chemistry, Massachusetts Institute of Technology, \mbox{77 Massachusetts Avenue, Cambridge MA 02139, USA}}
\author{Naoto Tsuji}
\affiliation{RIKEN Center for Emergent Matter Science (CEMS), Wako, Saitama 351-0198, Japan}
\author{Daisuke Hirai}
\affiliation{Department of Physics, University of Tokyo, Hongo, Tokyo 113-0033, Japan}
\author{Mircea Dinc\u{a}}
\affiliation{Department of Chemistry, Massachusetts Institute of Technology, \mbox{77 Massachusetts Avenue, Cambridge MA 02139, USA}}
\author{Hideo Aoki}
\affiliation{Department of Physics, University of Tokyo, Hongo, Tokyo 113-0033, Japan}

\date{\today}

\begin{abstract}
We design from first principles a new type of two-dimensional metal-organic frameworks (MOFs) using phenalenyl-based ligands to exhibit a half-filled flat band of the kagome lattice, which is one of the lattice family that shows Lieb-Mielke-Tasaki's flat-band ferromagnetism.
Among various MOFs, we find that \textit{trans}-Au-THTAP(trihydroxytriaminophenalenyl) has such an ideal band structure, where the Fermi energy is adjusted right at the flat band due to unpaired electrons of radical phenalenyl.
The spin-orbit coupling opens a band gap giving a non-zero Chern number to the nearly flat band, as confirmed by the presence of the edge states in first-principles calculations and by fitting to the tight-binding model.
This is a novel and realistic example of a system in which a nearly flat band is both ferromagnetic \textit{and} topologically non-trivial.
\end{abstract}


\maketitle

\textit{Introduction}. --- 
The exploration and discovery of new strongly-correlated or topologically non-trivial materials drive much of modern condensed-matter physics, yet an experimental design of such materials is still challenging. One class of materials that could greatly extend possibilities of material designing is, in our view, metal-organic frameworks (MOFs). These are crystalline materials composed of metal ions and bridging organic molecules, which have been the subject of numerous investigations in inorganic and materials chemistry~\footnote{MOF is defined as a coordination compound continuously extending in 1, 2 or 3 dimensions through coordination bonds with an open framework containing potential voids by IUPAC. See S. R. Batten, N. R. Champness, X.-M. Chen, J. Garcia-Martinez, S. Kitagawa, L. {\"O}hrstr{\"o}m, M. O'Keeffe, M. P. Suh, and J. Reedijk, Pure Appl. Chem. \textbf{85}, 1715 (2013).}. Owing to their typically trivial and localized electronic states, MOFs have not attracted much attentions from condensed-matter physicists. However, recent experimental success in fabricating atomically layered two-dimensional (2D) MOFs with kagome lattice structures, initiated by the Nishihara group~\cite{Kambe,Kambe2,Sakamoto,Dong}, is bridging the gap between condensed-matter physics and chemistry. The Dinc\u{a} group also succeeded in creating 2D MOFs~\cite{Sheberla,Campbell}. (Similar 2D MOFs have been fabricated by other groups~\cite{Shi,Yaghi,Jieshun,Huang}.) Some of these new 2D MOFs have been theoretically proposed to become organic $\mathbb Z_2$ topological insulators~\cite{Wang,Yang,SWYang,Zeng,Zhang,Dong2} or half-metallic ferromagnets~\cite{Zhao,Hu,Liu,ZhangSrep}.

The kagome lattice has a virtue of its electronic structure exhibiting a flat band at the highest (or lowest) energy. It has been proven that the tight-binding Hubbard model on the kagome lattice has a non-trivial ground state (far different from the atomic limit) showing itinerant ferromagnetism at arbitrary on-site Coulomb repulsion $U>0$ when the flat band is half-filled~\cite{Mielke}. While several classes of lattices are known to show the flat-band ferromagnetism, as proposed by Lieb, Mielke and Tasaki~\cite{Lieb,TasakiPRL,TasakiPTP,AokiMB}, the kagome lattice has an advantage that it is a realistic structure from a synthetic point of view, and does not require fine tuning for hopping parameters to accommodate a flat band~\footnote{Some honeycomb superlattice structures are also shown to have flat bands. See N. Shima and H. Aoki, Phys. Rev. Lett. \textbf{71}, 4389 (1993).}. When spin-orbit coupling (SOC) is introduced, the energy gap opens between neighboring bands, and the flat band becomes topologically non-trivial with a nonzero Chern number~\cite{Tang}. Although experimental results on a topological insulator in the bosonic system of the kagome lattice are known~\cite{TMI}, the fermionic electron system is more useful for possible applications in electronics and spintronics.
It has been discussed that if the nearly flat band with a non-zero Chern number is fractionally filled and is well separated from other bands, the system could be a fractional Chern insulator \cite{Tang,Neupert,Katsura}.

So far, it has been difficult to realize a partially filled flat band in a 2D kagome lattice (i.e., one where the Fermi energy is adjusted to the flat band). Whereas some quasi-one-dimensional (quasi-1D) inorganic systems~\cite{LBCO,Cecompound,Arita} and a quasi-1D organic doped molecular system~\cite{Garnica} have been proposed as possible candidates for nearly flat-band ferromagnetism, there is no experimental realization in 2D crystalline systems, as originally proposed by Lieb, Mielke, and Tasaki. Theoretically, Liu \textit{et al.}~\cite{LiuPRL} proposed that an In-based 2D organic nanosheet could possess a topologically non-trivial flat band near the Fermi energy; however, this material still needs additional hole doping to be ferromagnetic. Moreover, it may be difficult to experimentally realize such a 2D structure because the highly covalent bond between indium and carbon would not lead to the formation of crystalline materials.

In this Rapid Communication, we propose new 2D MOFs with kagome structures from first principles, where we can show that the flat band of a kagome lattice is indeed expected to be half-filled with the appropriate choice of organic ligand. The essential idea is to use an organic neutral radical called phenalenyl as a building block. In the absence of such an organic radical, hole doping would be necessary as described previously~\cite{LiuPRL,LiuCPB}. Based on first-principles electronic structure calculations, we discuss that the proposed phenalenyl-based 2D MOF becomes ferromagnetic with the flat band having a non-zero Chern number if SOC is taken into account.

\begin{figure}[t]
        \centering
        \includegraphics[width=8.6cm]{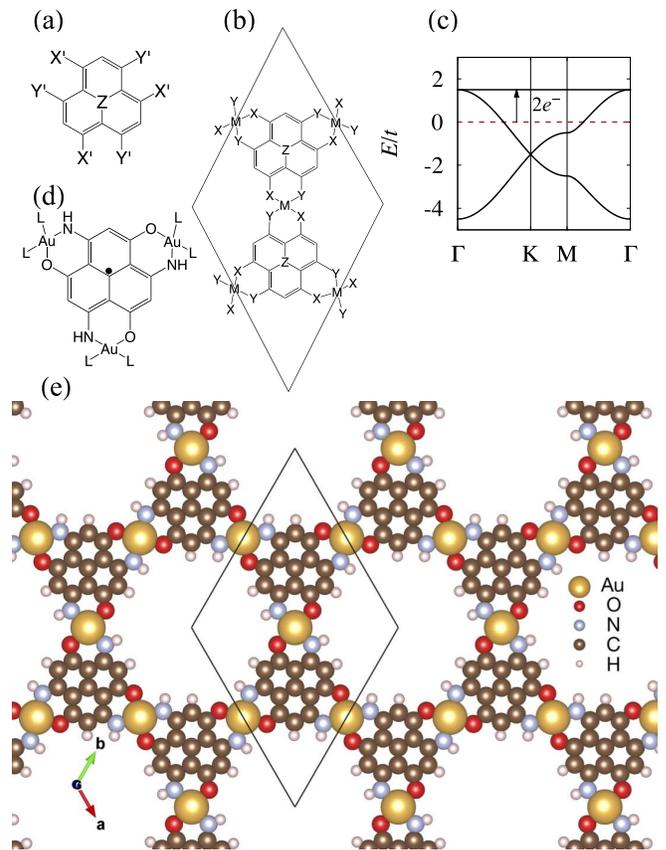}
\caption{(Color online) (a) Molecular structure of the phenalenyl-based ligand ($\mathrm{Z=C^{\bullet},N}$; $\mathrm{X^\prime,Y^\prime=OH,SH,NH_2}$).
(b) Structure of the unit cell (solid line) of the proposed MOFs ($\mathrm{M=Ni, Cu, Pt, Au}$; $\mathrm{X,Y=O, S, NH}$).
(c) Band structure of the single-orbital tight-binding model for the kagome lattice with a nearest-neighbor hopping parameter $t.$ The brown dashed line shows the Fermi energy when there is one electron per each orbital, while the arrow indicates that the flat band becomes half-filled when two unpaired electrons are transferred from radical phenalenyl.
(d) Structure of \textit{trans}-Au-THTAP, where $\mathrm{M=Au,\,X=NH,\,Y=O,\,Z=C^{\bullet}}$, and L stands for a neighboring ligand.
(e) Top view of \textit{trans}-Au-THTAP. The solid line indicates a unit cell.}
\label{mof}
\end{figure}

\textit{New 2D Metal-Organic Frameworks}. --- We employ phenalenyl-based ligands [see $\mathrm{Z=C^{\bullet}}$ in Fig.~\ref{mof}(a), where $\bullet$ represents an unpaired electron]. Owing to their triangular shape, these exhibit the appropriate symmetry to create kagome structures. These ligands are envisaged to be connected with a transition metal ($\mathrm{M=Ni, Cu, Pt, Au}$), where X and Y ($\mathrm{=O, S, NH}$) coordinate [see Fig.~\ref{mof}(b)]. If M is a spin-1/2 metal (Cu$^{2+}$ or Au$^{2+}$ in this case), there is one unpaired electron per orbital on each kagome site~\footnote{Here M = Ag is excluded since a homologous element Ag is known to have a tendency to form straight complexes.}. An important advantage of a phenalenyl-based ligand is that we can automatically adjust the Fermi energy from the brown dashed line in Fig.~\ref{mof}(c) to the desired position. This can result in half-filling of the flat band of the kagome lattice by transferring two unpaired electrons per unit cell to M due to the neutral stable radical resonance structure of phenalenyl [see the arrow in Fig.~\ref{mof}(c)]. In the absence of this radical (such as when $\mathrm{Z=N}$), electron or hole doping would be necessary, as described previously~\cite{Zhang,LiuPRL,LiuCPB}. The structure of \mbox{\textit{trans}-Au$_3$(THTAP)$_2$} (\mbox{\textit{trans}-Au-THTAP}, THTAP = trihydroxytriaminophenalenyl)~\footnote{An IUPAC name for the ligand is 3,6,9-triamino-phenalene-1,4,7-triol.} with $\mathrm{M=Au,\,X=NH,\,Y=O}$ is shown in Fig.~\ref{mof}(d). Its whole real-space structure and its unit cell (solid line) are illustrated in Fig.~\ref{mof}(e), with Au atoms occupying each vertex of the kagome lattice.

\textit{Electronic Structures From First Principles}. --- The above expectations to realize a half-filled flat band were confirmed from a first-principles electronic structure analysis. To this end, we used the first-principles electronic state calculation code called \textsc{openmx}~\cite{OpenMX}, based on density functional theory (DFT). With a repeated slab construction~\footnote{See Supplemental Material at [URL will be inserted by publisher] for the definition of the repeated slab construction.}, we first calculated possible phenalenyl-based MOFs with $\mathrm{M=Cu,\,Au}$ as spin-1/2 ions. In order to conserve a parity symmetry and break other symmetries to lift the degeneracy, we preferred $\mathrm{X\neq Y}$ and a \textit{trans}-structure~\cite{SWYang}~\footnote{A \textit{cis}-structure will break the structure's parity symmetry}. Then, we found that compounds with $\mathrm{M=Cu}$ tend to have a bended band, so we focused on $\mathrm{M=Au}.$  Finally, we found that all three remaining candidates (with $\mathrm{M=Au}$ and (X, Y) = (O, S), (S, NH), (NH, O)) have a nearly flat band exactly lying on the Fermi energy \footnote{See Supplemental Material at [URL will be inserted by publisher] for all the calculated band structures.}. Among these, \mbox{\textit{trans}-Au-THTAP} [(X,Y)=(NH, O)] has the optimal band structure in the sense that its band structure is accurately matched to that obtained from a tight-binding model on the kagome lattice around the Fermi energy.

\begin{figure*}
        \centering
        \includegraphics[width=17.8cm]{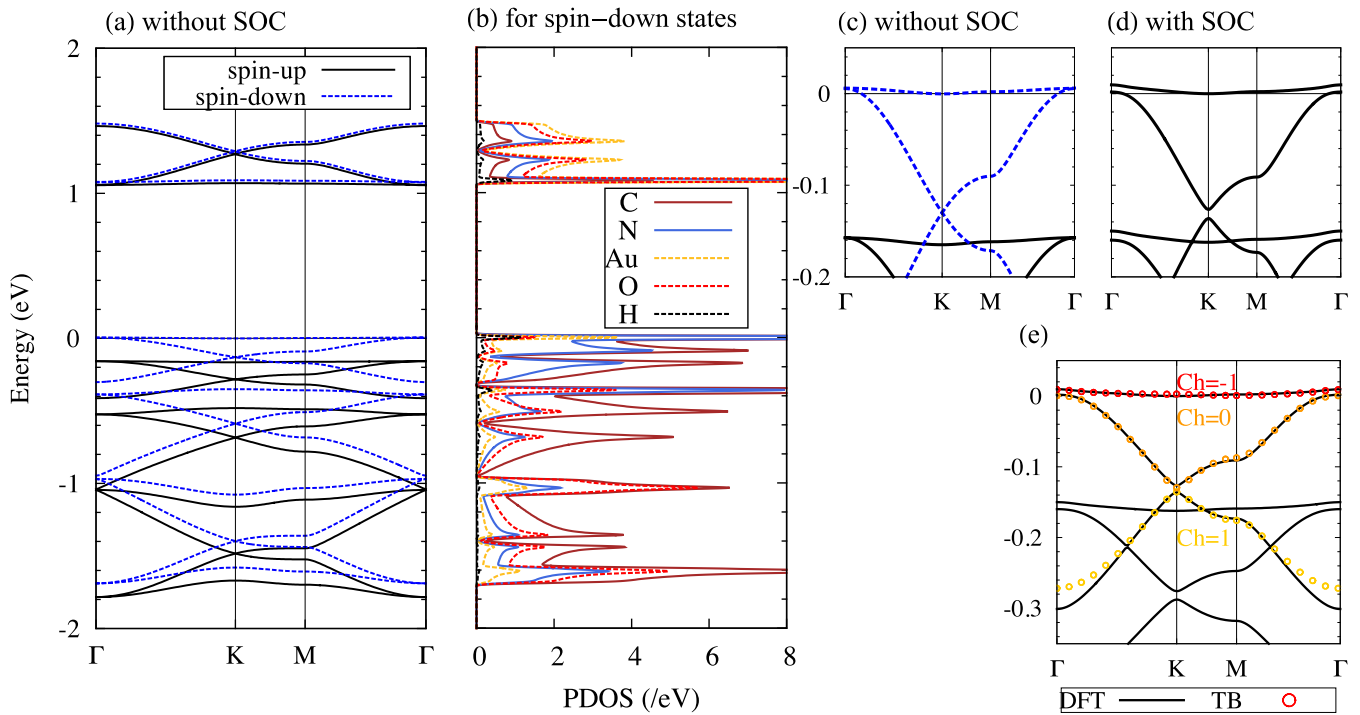}
        \caption{(Color online) (a) Band structure of \textit{trans}-Au-THTAP calculated without SOC. The black solid (blue dashed) lines show the spin-up (spin-down) bands. (b) PDOS of the spin-down bands for each element in \textit{trans}-Au-THTAP calculated without SOC.  (c) Zoom-in of (a) around the Fermi energy. (d) Zoom-in of the band structure calculated with SOC. (e) Comparison between the band structures calculated with SOC by DFT (solid lines) and by TB (circles) with the calculated Chern numbers (Ch). In each panel, the Fermi energy is taken to be zero.}
        \label{band}
\end{figure*}

\textit{trans-Au-THTAP}. --- After geometric optimization, \textit{trans}-Au-THTAP was revealed to favor a planar structure~\footnote{See the cif files in Supplemental Material at [URL will be inserted by publisher] for the relaxed atomic structures.} with an optimized lattice constant of 16.91 \AA. The band structure calculated without SOC is shown in Fig.~\ref{band}(a). The black solid lines display the spin-up bands, while the blue dashed lines display the spin-down bands. The system shows a metallic nature and the nearly flat band near the Fermi energy [$E=0$ in Fig.~\ref{band}(a)] arising from the kagome lattice is approximately half-filled and indeed spin-polarized. This gives a ferromagnetic behavior with a total spin moment of 1.00 $\mathrm{\mu_B/unit\,cell}$. We have to note that this spin moment exactly coincides with the expected value for flat-band ferromagnetism~\cite{Mielke}. Remarkably, the analysis of the partial density of states (PDOS) for each element clearly shows that the kagome bands near the Fermi energy mostly come from C and N atoms and less from Au $d$-orbitals [see Fig.~\ref{band}(b)]. This real-space property is further confirmed by the analysis of spin density~\footnote{See Supplemental Material at [URL will be inserted by publisher] for the real-space spin density and Bloch functions.} and could be explained by the itinerant mechanism of ferromagnetism rather than by the interacting localized moments on Au. Moreover, by carrying out a fully relativistic self-consistent calculation on this system, we find that SOC opens a gap of 7.8 meV between the nearly flat band and the lower dispersive band at $\Gamma$ [compare Fig.~\ref{band}(c) without SOC to (d) with SOC]. This system has turned out to be still metallic, due to a slight warping of the nearly flat band. The spin-up and spin-down bands are no longer separable when we calculate with SOC, but the $z$-component of the spin ($\sigma^z$) is approximately a conserved quantum number because the calculated magnetic order is always along the $z$-direction in the case of including SOC. This gives a total spin moment of 0.99 $\mathrm{\mu_B/unit\,cell},$ a total orbital moment of \mbox{0.02 $\mathrm{\mu_B/unit\,cell}$} along the $z$-direction, and an exchange splitting of \mbox{159.5 meV}.

\textit{Topological Properties From a Tight-Binding Model}. --- In order to show the topological non-triviality of the gap between the nearly flat band and the dispersive band of \textit{trans}-Au-THTAP, we first considered a single-orbital tight-binding (TB) model on the kagome lattice, where each single orbital is assumed to be localized around Au. Actually, the wave functions forming the kagome bands are not completely localized on Au and spreading over the $\pi$-conjugated system, but we can still assume a single-orbital TB model as long as the lattice symmetry is preserved and the parameters of the TB model are somehow renormalized by the effect of spreading. We added a Zeeman term (exchange splitting) to the Hamiltonian considered in~\cite{Wang,Yang,SWYang,Franz,Tang} to include the effect of ferromagnetism~\cite{Zhang}. We considered a complex nearest-neighbor (NN) hopping and a real next-nearest neighbor (NNN) hopping in a Hamiltonian, $H= E_0+H_0+H_{\textrm{SO}}+H_{\textrm{Z}},$ where
\begin{align}
        H_0 &= -t_1 \sum_{\langle ij \rangle\sigma}c^\dagger_{i\sigma}c_{j\sigma}-t_2 \sum_{\langle\!\langle ij \rangle\!\rangle\sigma}c^\dagger_{i\sigma}c_{j\sigma}, \\
        H_{\textrm{SO}} &= i\lambda_1 \sum_{\langle ij \rangle\alpha\beta} \nu_{ij}\sigma^z_{\alpha\beta}c^\dagger_{i\alpha}c_{j\beta}, \\
        H_{\textrm{Z}} &= b\sum_i (-1-\sigma^z_{\alpha\beta}) c^\dagger_{i\alpha}c_{i\beta}.
\end{align}
Here $E_0$ is the energy offset and $c_{i\sigma}^\dagger$ and $c_{i\sigma}$ are the creation and annihilation operators of the $\sigma$-spin electron on the $i$th site of the kagome lattice, respectively. $\langle ij \rangle$ and $\langle\!\langle ij \rangle\!\rangle$ denote the NN  and NNN bonds, respectively, while $t_1$ and $t_2$ are the corresponding real-valued hopping parameters. $\lambda_1$ is the NN intrinsic spin-orbit coupling and $b$ is the Zeeman splitting, while $\bm{\sigma}_{\alpha\beta}$ is the Pauli matrix for the spin component. $\nu_{ij}$ is 1 for the counterclockwise hopping and $-1$ for the clockwise hopping when viewed from above. For simplicity, we have omitted the NNN imaginary hopping parameter $\lambda_2,$ which is expected to be much smaller than the others~\cite{Yang}.

This TB Hamiltonian conserves the $z$-component of the spin, so we can divide the one-particle Hilbert space $\mathcal{H}$ into $\mathcal{H_\uparrow}\oplus \mathcal{H_\downarrow}$ by the eigenvalue of $\sigma^z.$ We only consider the space $\mathcal{H_\downarrow}$ because there are only spin-down bands near the Fermi energy. In other words, we have projected out the spin-up states by taking the limit $b \to\infty$ first. We can then accurately fit the kagome bands in the DFT calculation with SOC (solid line) to that obtained from the TB model (circles) around the Fermi energy as shown in Fig.~\ref{band}(e) with the parameters $E_0=-87.4\,\textrm{meV},\,t_1=45.1\,\textrm{meV},\,t_2=1.0\,\textrm{meV},\,\lambda_1=1.2\,\textrm{meV}.$ Based on the TB model, we can calculate a topological Chern number (Ch) for each band~\cite{FHS,HFA06}. From the results displayed in Fig.~\ref{band}(e), we can conclude that the nearly flat band is indeed topologically non-trivial with a Chern number of $-1$ within this TB framework.

\begin{figure}
\centering
\includegraphics[clip, width=8.6cm]{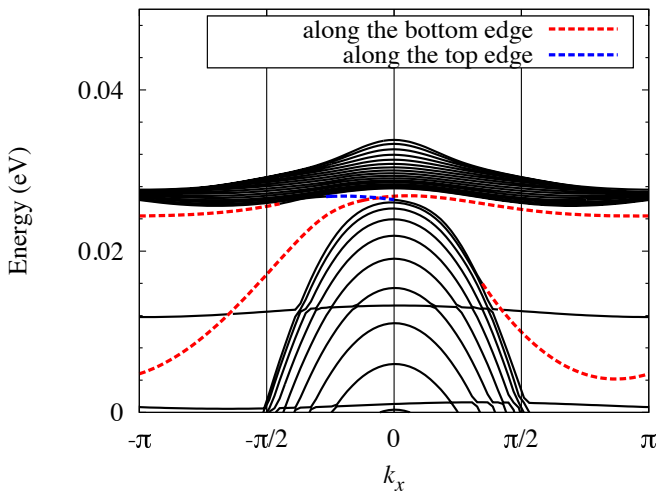}
\caption{(Color online) Band structure for a ribbon of \textit{trans}-Au-THTAP. The black solid lines represent the bulk states, and the red (blue) dashed lines represent topological edge states along the bottom (top) edge.}
\label{edge}
\end{figure}

\textit{Edge States}. --- Because the topological property of the SOC gap is model-dependent, the choice of the SOC term in the TB model can be somewhat arbitrary. A clear way to show the topological non-triviality of the system is to detect the topological edge states by a DFT calculation. To do so, we again used \textsc{openmx} with a repeated ribbon construction~\footnote{See Supplemental Material at [URL will be inserted by publisher] for the definition of the repeated ribbon construction.}.
The calculated band structure is shown in Fig.~\ref{edge}, where the red and blue dashed lines clearly show the chiral edge states for each boundary and cannot be gapped away. Since we took a different boundary condition for each edge (one with Au aligning along the boundary and the other without Au), they are asymmetric against $k_x=0.$ These edge modes are similar to the ones in Ref.~\onlinecite{SWYang} except for the spin-polarization due to its ferromagnetism.
The emergence of these non-trivial states localized along the boundaries~\footnote{See Supplemental Material at [URL will be inserted by publisher] for the real-space non-trivial properties of the edge states.} again confirms the topological nature of the nearly flat band. These two results from TB and DFT show that there exists an exotic phase --- a topologically non-trivial nearly flat band with full spin polarization.

\textit{Phonon Calculations}. --- To verify the stability of the proposed geometric structure, we also performed phonon calculations. We used the first-principles electronic state calculation code called \textsc{Quantum ESPRESSO}~\cite{QE-2009}. These studies show that a flat, free-standing sheet of \textit{trans}-Au-THTAP may buckle at low temperature~\footnote{See Supplemental Material at [URL will be inserted by publisher] for the phonon dispersion.}, giving rise to imaginary out-of-plane phonon modes [see Fig.~S5 in the Supplemental Material]. Experimentally, however, sheets of the target material would likely be tested on a flat surface, not free-standing. Thus, we expect that the out-of-plane modes may be suppressed under experimental conditions, thereby retaining the non-trivial properties. Proximity effects, such as Rashba-type SOC, from various substrates on the electronic and magnetic structure of this material could itself make the subject of interesting future theoretical studies.

\textit{Conclusion}. --- We have proposed several new 2D MOFs, and found that \textit{trans}-Au-THTAP has a topologically non-trivial nearly flat band from the DFT calculations. This is a novel and realistic example of a system in which a nearly flat band at the Fermi energy is both ferromagnetic \textit{and} topologically non-trivial. From a synthetic standpoint, we have to note that Au indeed prefers a square planar coordination environment in the +2 formal oxidation state, as discussed in Ref.~\onlinecite{Au}. We also note that there would be electron correlation effects for the flat-band ferromagnetic ground state that are not properly captured within the DFT framework, which is worthwhile to study in the future.

Although the proposed system does not have a quantized Hall current due to its metallicity, a topologically non-trivial phase realized in the proposed material (called a Chern metallic phase in Ref.~\onlinecite{CMP}) is still worth investigating. At this moment we have only confirmed the existence of the edge states with a constraint on the spin structure. Future work will probe whether magnetic order can exist along the boundaries and how it would affect the topological edge states. Furthermore, it would be interesting to explore whether one could enhance the band gap given the known tunability of MOFs. This would create a ferromagnetic insulator with a quantized Hall conductance (quantum anomalous Hall effect~\cite{QAHE}) or a fractional Chern insulator~\cite{Tang,Neupert,Katsura,Bergholtz,FCI,FTI,FTI2,FQHE} with a large (band gap)/(band width) ratio. We note, in this sense, that a no-go theorem has been proven mathematically for topologically non-trivial perfectly flat bands within local tight-binding models~\cite{NGT}.

\begin{acknowledgements}
We wish to acknowledge H. Nishihara, R. Sakamoto and N. Shima for illuminating discussions in the early stage of the present work, and M. Oshikawa, S. Kasamatsu, M. Kawamura, S. Tsuneyuki, G. Jackeli, H. Takagi, T. Ozaki, R. Osawa, A. Pulkin and Y. Murotani for helpful comments. This work was supported in part by MEXT Grant Nos. 26247064, 25107005 and 25610101, and by ImPACT (No. 2015-PM12-05-01) from JST (HA). M.G.Y. is supported by the Materials Education program for the future leaders in Research, Industry, and Technology (MERIT). N.T. is supported by Grants-in-Aid for Scientific Research from JSPS (Nos. 25104709 and 25800192). M.D. is supported by the Center for Excitonics, an Energy Frontier Research Center funded by the U.S. Department of Energy, Office of Science, Office of Basic Energy Sciences under award number DE-SC0001088 (MIT), and by non-tenured faculty awards from the Sloan Foundation, the Research Corporation for Science Advancement (Cottrell Scholars), and 3M. The computation in this work has been done with the facilities of the Supercomputer Center, the Institute for Solid State Physics, the University of Tokyo. The initial crystal structures for the geometry optimization have been taken from the Cambridge Crystallographic Data Centre.
\end{acknowledgements}

\bibliography{paper}

\end{document}

\title{Supplemental Material for ``First-Principles Design of a Half-Filled Flat Band \\ of the Kagome Lattice in Two-Dimensional Metal-Organic Frameworks''}
\author{Masahiko G. Yamada}
\affiliation{Institute for Solid State Physics, University of Tokyo, Kashiwa 277-8581, Japan}
\affiliation{Department of Physics, University of Tokyo, Hongo, Tokyo 113-0033, Japan}
\author{Tomohiro Soejima}
\affiliation{Department of Chemistry, Massachusetts Institute of Technology, \mbox{77 Massachusetts Avenue, Cambridge MA 02139, USA}}
\author{Naoto Tsuji}
\affiliation{RIKEN Center for Emergent Matter Science (CEMS), Wako, Saitama 351-0198, Japan}
\author{Daisuke Hirai}
\affiliation{Department of Physics, University of Tokyo, Hongo, Tokyo 113-0033, Japan}
\author{Mircea Dinc\u{a}}
\affiliation{Department of Chemistry, Massachusetts Institute of Technology, \mbox{77 Massachusetts Avenue, Cambridge MA 02139, USA}}
\author{Hideo Aoki}
\affiliation{Department of Physics, University of Tokyo, Hongo, Tokyo 113-0033, Japan}

\maketitle

\onecolumngrid

Section A: calculation (simulation) details of the band structures and the edge states of the proposed materials within the framework of density functional theory (DFT). Section B: calculated band structures for all the proposed materials explaining the emphasis on \mbox{$\textit{trans}$-Au-THTAP(trihydroxytriaminophenalenyl)} as a target metal-organic framework (MOF). Section C: discussion on the exchange-correlation-potential dependence of the calculated band structures of $\textit{trans}$-Au-THTAP. Section D: bulk spin density and bulk Bloch functions of $\textit{trans}$-Au-THTAP. Section E: wave functions of the edge states of $\textit{trans}$-Au-THTAP and distinction of the topological edge states from the trivial ones. Section F: calculated phonon dispersion for $\textit{trans}$-Au-THTAP.

\section*{S\lowercase{ection} A: Profiles for the DFT calculations}
In \textsc{openmx}~\cite{OpenMX}, one-particle wave functions are expressed by the linear combination of pseudoatomic basis functions, where the norm-conserving pseudopotentials are used. In our simulation, the generalized gradient approximation represented by Perdew, Burke and Ernzerhof (GGA-PBE)~\cite{PBE} was employed for the exchange-correlation potential. A self-consistent loop was iterated until the energy was relaxed with the error of $10^{-7}$ Hartree. The geometric (structure) optimization of internal coordinates was also iterated until the force became smaller than $10^{-4}$ Hartree/Bohr, and the lattice constant was optimized with the error of \mbox{0.01 \AA}. The plot of the density of states (in Fig.~2(b) in the main text) used a Gaussian broadening of 0.01 eV. Initial crystal structures for the geometry optimization were constructed from the molecular structures \footnote{CCDC structure codes: XIGZAQ and MEYZUN.} in the Cambridge Crystallographic Data Centre~\cite{CCDC,CSD}.

In order to capture the two-dimensional (2D) nature of the proposed MOFs, we defined a repeated slab construction as follows. The distance between the neighboring MOFs is 10 \AA\hspace{0.1em} to neglect the interlayer interaction. We used a $30\times 30\times 1$ $k$-points mesh in the first Brillouin zone and an energy cutoff of 500 Ry for the numerical integrations and the solution of the Poisson equation using the fast Fourier transformation algorithm. The semicore 5$p$ electrons of Au and the semicore 3$s$ and 3$p$ electrons of Cu were treated as valence electrons. As basis functions, 5$p$, 5$d$, 5$f$, 6$s$, 6$p$, 6$d$, 7$s$, 7$p$, and 8$s$ orbitals for Au, 3$s$, 3$p$, 3$d$, 4$s$, 4$p$, 4$d$, 4$f$, 5$s$, and 5$p$ orbitals for Cu, 2$s$, 2$p$, 3$s$, 3$p$, and 3$d$ orbitals for C, N, O, 3$s$, 3$p$, 3$d$, 4$s$, and 4$p$ orbitals for S, and 1$s$, 2$s$, 2$p$, and 3$p$ orbitals for H were employed. We have checked the convergence of the total energy against the increase in the number of basis functions, and we also reproduced the band structure and magnetization using the plane wave code, \textsc{Quantum ESPRESSO}~\cite{QE-2009}. Therefore, the choice of the basis functions does not affect our conclusion. The calculation details for \textsc{Quantum ESPRESSO} are included in Section~F.

In order to calculate the edge states of \textit{trans}-Au-THTAP using \textsc{openmx}, we defined a repeated ribbon construction as follows. We considered a ribbon of the proposed MOF with the width of 292.8 \AA\hspace{0.1em} (which corresponds to 20 unit cells) along the $y$-direction [i.e., perpendicular to both the $z$-direction and the $a$-direction. See Fig.~S\ref{secb}(g)]. We used a $30\times 1\times 1$ $k$-points mesh. We did not execute the optimization of the atomic configuration and the lattice constant for simplicity. Because we are interested in the topological properties of the proposed bulk structure, we assumed the magnetic easy axis to be the same as the bulk, and imposed the spin orientation on the constraint to be along the $z$-direction, which is taken to suppress the boundary effect that destabilizes the bulk magnetic ordering.

\newpage

\section*{S\lowercase{ection} B: Calculated band structures for all the proposed materials}

\begin{table}
        \centering
        \caption{\label{lc}Lattice constants and total spin moments obtained from the geometric optimization and the DFT calculation with spin polarization for the proposed materials.}
        \begin{ruledtabular}
        \begin{tabular}{ccccccc}
                & Name & Ligand & M & (X, Y) & Lattice constant (\AA) & Total spin moment ($\mathrm{\mu_B/unit\,cell}$)\\
                \hline
                (a) & \textit{trans}-Cu-THTTP & trihydroxytrithiophenalenyl & Cu & (O, S) & 17.29 & 1.00 \\
                (b) & \textit{trans}-Cu-TATTP & triaminotrithiophenalenyl & Cu & (S, NH) & 17.50 & 1.00 \\
                (c) & \textit{trans}-Cu-THTAP & trihydroxytriaminophenalenyl & Cu & (NH, O) & 16.62 & 4.80 \\
                (d) & \textit{trans}-Au-THTTP & trihydroxytrithiophenalenyl & Au & (O, S) & 17.70 & 1.00 \\
                (e) & \textit{trans}-Au-TATTP & triaminotrithiophenalenyl & Au & (S, NH) & 17.90 & 1.00 \\
                (f) & \textit{trans}-Au-THTAP & trihydroxytriaminophenalenyl & Au & (NH, O) & 16.91 & 1.00
        \end{tabular}
\end{ruledtabular}
\end{table}

\begin{figure}
        \centering
        \includegraphics[width=17.8cm]{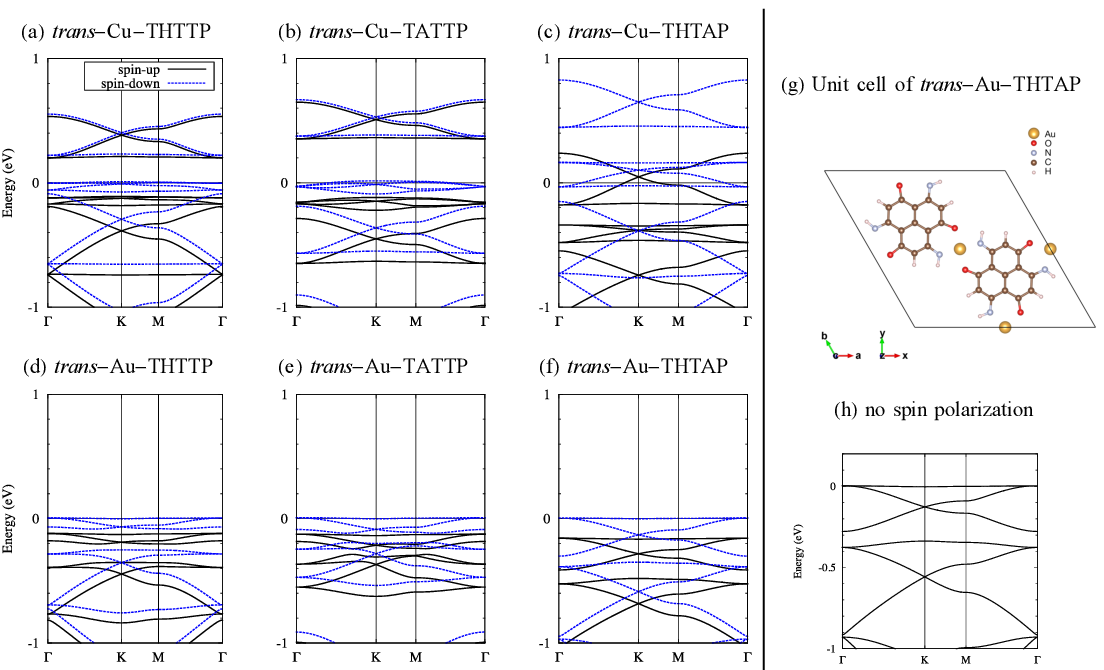}
        \caption{(a)-(f) Band structures for (a) \textit{trans}-Cu-THTTP, (b) \textit{trans}-Cu-TATTP, (c) \textit{trans}-Cu-THTAP, (d) \textit{trans}-Au-THTTP, (e) \textit{trans}-Au-TATTP and (f) \textit{trans}-Au-THTAP. In each panel, black solid (blue dashed) lines show the spin-up (spin-down) bands, and the Fermi energy is taken to be zero. (g) Unit cell of \textit{trans}-Au-THTAP and the definition of the axes. (h) Band structure for \textit{trans}-Au-THTAP calculated without spin polarization.}
        \label{secb}
\end{figure}

With the repeated slab construction, we have calculated all the proposed materials without a spin-orbit coupling (SOC) and found that all of these favor a planar hexagonal structure and ferromagnetism. The calculated lattice constants and total spin moments are listed in Table~\ref{lc}, and the calculated band structures are shown in Figs.~S\ref{secb}(a)-(f). It can be easily seen that (f) \textit{trans}-Au-THTAP has the best band structure in the sense that its band structure is accurately fitted to that obtained from a tight-binding model on the kagome lattice around the Fermi energy, as shown in Fig.~2(e) in the main text.
In (d) \textit{trans}-Au-THTTP and (e) \textit{trans}-Au-TATTP, the lowest band of the kagome bands near the Fermi energy is warping around $\Gamma$ due to the interference with the lower irrelevant bands, and these bands will not be described by a simple single-orbital kagome tight-binding model. Before introducing spin polarization, the flat band of the kagome lattice is completely half-filled in \textit{trans}-Au-THTAP [see Fig.~S\ref{secb}(h)]. This unpolarized state will be destabilized and a ferromagnetic order appears in the calculation with spin polarization [see Fig.~S\ref{secb}(f)].

\newpage

\section*{S\lowercase{ection} C: Comparison between GGA and GGA+U for \lowercase{\textit{trans}}-A\lowercase{u}-THTAP}

\begin{figure}
        \centering
        \includegraphics[width=16cm]{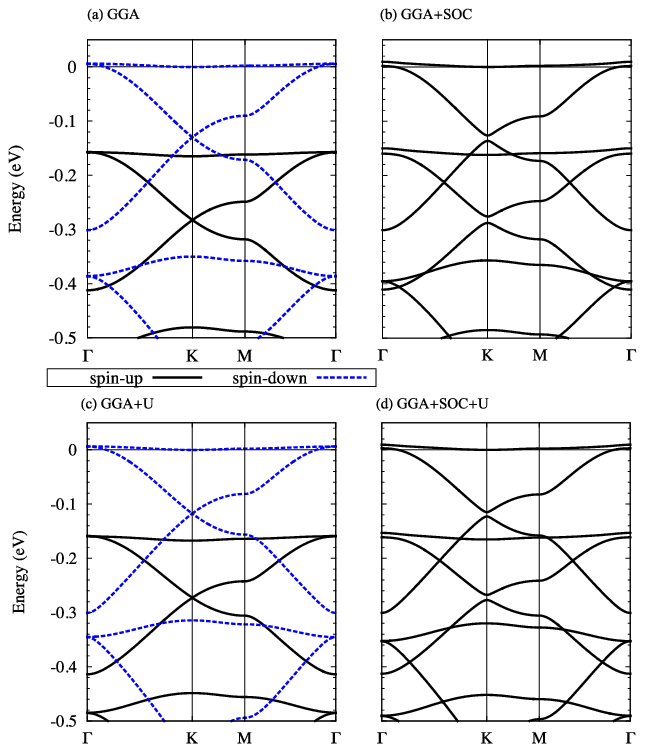}
        \caption{Band structures for \textit{trans}-Au-THTAP in (a) GGA, (b) GGA+SOC, (c) GGA+U and (d) GGA+SOC+U calculations. In calculations without SOC, black solid (blue dashed) lines show the spin-up (spin-down) bands. In each panel, the Fermi energy is taken to be zero.}
        \label{secc}
\end{figure}

With the repeated slab construction for \textit{trans}-Au-THTAP, we have compared the band structures of \textit{trans}-Au-THTAP between GGA(+SOC) and GGA(+SOC)+U, where +SOC means calculations with a spin-orbit coupling. In GGA(+SOC)+U calculations, we added a Hubbard term of $U=2$ eV for Au 5$d$ orbitals (which is a value used often for 5$d$ transition metals, such as Ir~\cite{Ir}). As shown in Fig.~S\ref{secc}, there is no qualitative difference between GGA and GGA+U for the kagome bands, but by including $U$ the exchange splitting grows from 159.5 meV to 162.5 meV, and the SOC gap between the flat band around the Fermi energy and the lower dispersive band changes from 7.8 meV to 6.5 meV. This property agrees with the fact that the flat-band ferromagnetism appears for any $U>0$~\cite{Mielke}, and the effect of the (at least infinitesimal) correlation is already included in GGA(+SOC) calculations. It must be noted that van der Waals (vdW) forces are expected to have smaller effects on the band structure because the atoms in the proposed material are connected by covalent or coordination bonds rather than by vdW forces, and the calculation including vdW forces (the DFT-D2 method) does not alter the band structure in a meaningful way.

\newpage

\section*{S\lowercase{ection} D: Bulk spin density and bulk Bloch functions of \lowercase{\textit{trans}}-A\lowercase{u}-THTAP}

\begin{figure}
        \centering
        \includegraphics[width=17.8cm]{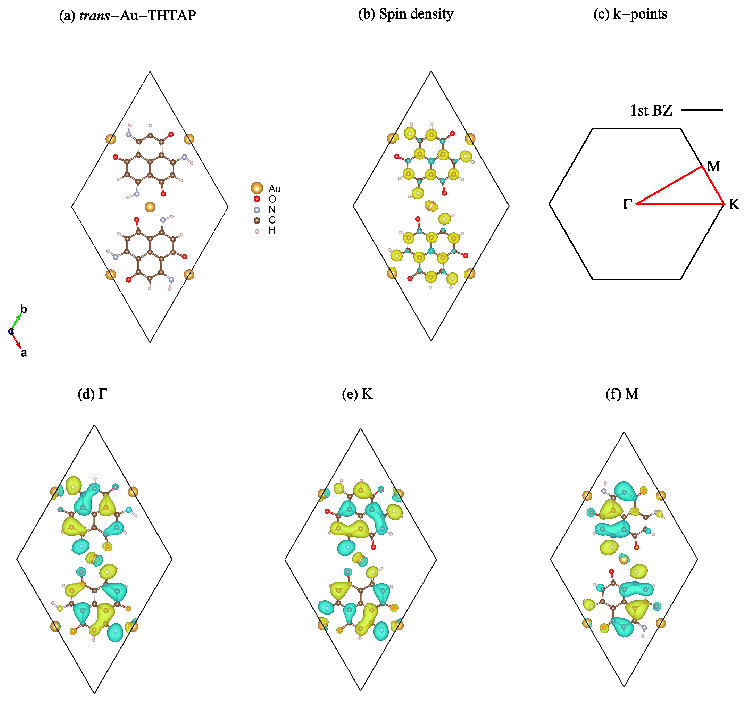}
        \caption{(a) Unit cell of \textit{trans}-Au-THTAP.
(b) Isosurfaces for the spin density of \textit{trans}-Au-THTAP, where yellow and blue bubbles represent isosurfaces with plus and minus signs, respectively.
(c) First Brillouin zone (1st BZ) shown in black solid line. Typical $k$-points are also plotted.
(d)-(f) Isosurfaces for the real parts of the spin-down Bloch functions for (d) $\Gamma$, (e) K and (f) M points in the 1st BZ.}
        \label{secd}
\end{figure}

With the repeated slab construction for \textit{trans}-Au-THTAP, we calculated the bulk spin density without SOC [see Fig.~S\ref{secd}(b)]. In addition to the partial density of states (PDOS) analysis shown in Fig.~2(b) in the main text, the spin density clearly shows the real-space profile of the flat band because almost only the flat band is spin-polarized in the ground state of \textit{trans}-Au-THTAP. It can be clearly seen that the flat band comes mostly from C 2$p_z$ and N 2$p_z$ orbitals, with a smaller contribution from Au 5$d$ orbitals. We also calculated the real parts of the spin-down Bloch functions (the periodic part of the one-particle wave function for each $k$-point) for the flat band near the Fermi energy with SOC [see Figs.~S\ref{secd}(d)-(f)]. There is a strong momentum ($k$) dependence in the Bloch functions, which is a feature inherent in the nearly-flat-band ferromagnetism~\cite{kdep}, clearly shows that the flat band is not in the trivial atomic limit.

\newpage

\section*{S\lowercase{ection} E: Real-space properties of the edge states of \lowercase{\textit{trans}}-A\lowercase{u}-THTAP}

\begin{figure}
        \centering
        \includegraphics[width=17.8cm]{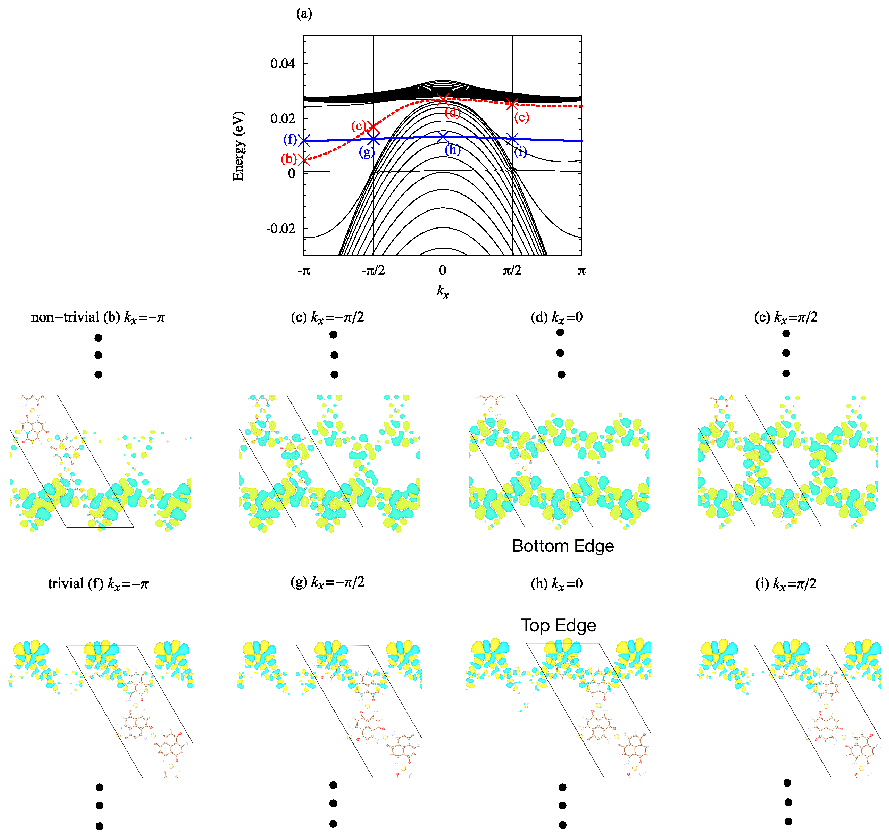}
        \caption{(a) Band structure for a ribbon of \textit{trans}-Au-THTAP. We highlighted the non-trivial edge state along the bottom edge (red dashed line) and the trivial edge state along the top edge (blue solid line). (b)-(e) Real parts of spin-down Bloch functions for the non-trivial edge state along the bottom edge at (b) $k_x=-\pi,$ (c) $k_x=-\pi/2,$ (d) $k_x=0$ and (e) $k_x=\pi/2.$ (f)-(i) Real parts of typical spin-down Bloch functions for the trivial edge state along the top edge at (f) $k_x=-\pi,$ (g) $k_x=-\pi/2,$ (h) $k_x=0$ and (i) $k_x=\pi/2.$}
        \label{sece}
\end{figure}

When we calculate the edge states of \textit{trans}-Au-THTAP with the repeated ribbon construction, there could be both non-trivial edge states from the topological property and trivial edge states unrelated to the topological property of the bulk. If we exclude such trivial bands as states in the atomic limit, the existence of the non-trivial edge states shows the topological nature of the nearly flat band, but in order to distinguish the non-trivial states from the trivial ones we have to check the properties of the real-space wave functions. The wave functions for the dashed bands shown in Fig. 3 in the main text are exponentially localized along the boundaries, while there are also trivial flat bands localized along the boundaries. To clarify the difference using concrete examples, we highlighted the non-trivial edge state along the bottom edge (red dashed line) and the trivial edge state along the top edge (blue solid line) in the band structure of the edge states shown in Fig.~S\ref{sece}(a), which is the same band structure as in the main text [see Fig.~3 in the main text]. Figures~S\ref{sece}(b)-(e) display real parts of typical spin-down Bloch functions for the non-trivial edge state at each $k$-point~\cite{Suwa}, while (f)-(i) for the trivial edge state. There is a significant momentum ($k_x$) dependence in the Bloch functions for the non-trivial edge state, whereas for the trivial edge state the real-space structure is almost the same for different $k_x$. Such a difference in the momentum dependence was checked for every edge state. Thus, we can conclude that the trivial edge state is actually in the atomic limit, so we can exclude such states when we consider the topological property of the proposed system.

\section*{S\lowercase{ection} F: calculated phonon dispersion for \lowercase{\textit{trans}}-A\lowercase{u}-THTAP}

\begin{figure}
        \centering
        \includegraphics[width=12cm]{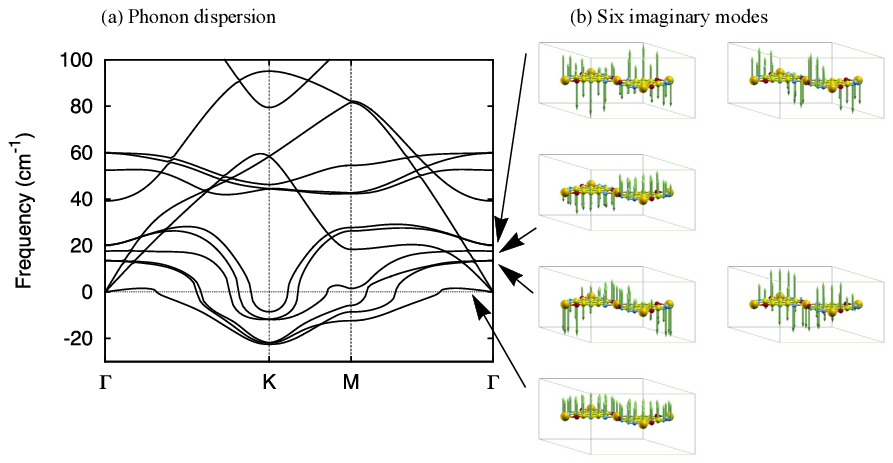}
        \caption{(a) Phonon dispersion for the planar structure of \textit{trans}-Au-THTAP. In this figure, imaginary frequencies $\omega$ are actually shown by negative values $-|\omega|.$ (b) Force vectors shown as green arrows for the six out-of-plane modes at $\Gamma,$ which are adiabatically connected to the six imaginary modes at K.}
        \label{secf}
\end{figure}

To calculate the phonon dispersion, we used the first-principles electronic state calculation code called \textsc{Quantum ESPRESSO}~\cite{QE-2009}. In \textsc{Quantum ESPRESSO}, one-particle wave functions are expressed by the linear combination of plane waves and the ultrasoft pseudopotentials taken from the THEOS website~\cite{THEOS} are used. In this simulation, GGA-PBE~\cite{PBE} was again employed for the exchange-correlation potential. A self-consistent loop was iterated until the energy was relaxed with the error of $10^{-16}$ Ry. Assuming the completely planar structure with a P6/m symmetry, the geometric optimization of internal coordinates was also iterated until the force became smaller than $10^{-4}$ Ry/a.u. and the lattice constant was also relaxed to 16.88 \AA. We used the repeated-slab construction for the free-standing \textit{trans}-Au-THTAP with $6\times 6\times 1$ $k$-points mesh in the first Brillouin zone for the electronic state calculations and $3\times 3\times 1$ $q$-points mesh for the phonon calculations. We took energy cutoffs of 50 Ry for wavefunctions and 500 Ry for electron densities.

After applying an acoustic sum rule, which reduces a numerical artifact of the imaginary phonons around $\Gamma,$ we get the phonon dispersion shown in Fig.~S\ref{secf}(a). In this figure, imaginary frequencies $\omega$ are actually displayed by negative values $-|\omega|.$ There appear six imaginary modes around K, which suggests structural instabilities in a free-standing slab of \textit{trans}-Au-THTAP with a P6/m symmetry. However, these modes are all out-of-plane modes along the $z$-direction, which can be seen from the force vectors at $\Gamma$ shown in Fig.~S\ref{secf}(b). For the other $q$-points, the force vectors are complex numbers and cannot easily be drawn in the figure, but we have checked that the complex force vectors only have $z$-components. Therefore, we expect that by supporting a flat sheet of \textit{trans}-Au-THTAP on a flat substrate, which is a more realistic experimental scenario than the free-standing sheet used in our calculations, these modes will be suppressed.

\newpage

\bibliography{suppl}